\documentclass[doublespacing]{elsart}

\usepackage{epsfig}
\usepackage{graphics}
\usepackage{amsmath}
\usepackage{amssymb}
\usepackage{hhline}
\usepackage[tight]{subfigure}
\usepackage{here}

\newcommand{\reaktion}{\mbox{$pp\to \,pp\:\!K^+\!K^- $ }}

\newif\ifdraft\draftfalse

\newcommand{\notiz}[1]{%
\ifdraft {\marginpar{\small \bf #1}}%
\fi
}

\newif\ifbw\bwfalse

\begin{document}
\begin{frontmatter}

\title{Kaon pair production close to threshold}
\author[1]{P.~Winter}\ead{winter@npl.uiuc.edu},
\author[1]{M.~Wolke},    
\author[2]{H.-H.~Adam},
\author[3]{A.~Budzanowski},
\author[4]{R.~Czy\.zykiewicz},
\author[1]{D.~Grzonka},
\author[4]{M.~Janusz},
\author[4]{L.~Jarczyk},
\author[4]{B.~Kamys},
\author[2]{A.~Khoukaz}, 
\author[1]{K.~Kilian},
\author[4]{P.~Klaja},
\author[1,4]{P.~Moskal},
\author[1]{W.~Oelert},
\author[4]{C.~Piskor-Ignatowicz},
\author[4]{J.~Przerwa},
\author[1]{J.~Ritman},
\author[1,5]{T.~Ro\.zek},
\author[1]{T.~Sefzick},
\author[5]{M.~Siemaszko}, 
\author[4]{J.~Smyrski}, 
\author[2]{A.~T\"aschner},
\author[6]{P.~W{\"u}stner},
\author[1]{Z.~Zhang},
\author[5]{W.~Zipper}         

\address[1]{Institut f{\"u}r Kernphysik, Forschungszentrum J\"{u}lich, D-52425 J\"ulich, Germany}
\address[2]{Institut f{\"u}r Kernphysik, Westf{\"a}lische Wilhelms--Universit{\"a}t,  D-48149 M{\"u}nster, Germany}
\address[3]{Institute of Nuclear Physics, PL-31-342 Cracow, Poland}
\address[4]{Institute of Physics, Jagellonian University, PL-30-059 Cracow, Poland}
\address[5]{Institute of Physics, University of Silesia, PL-40-007 Katowice, Poland}
\address[6]{Zentrallabor f{\"u}r Elektronik,  Forschungszentrum J\"{u}lich, D-52425 J\"ulich, Germany}

\begin{abstract}
The total cross section of the reaction \reaktion has been measured at excess energies $Q=10$\,MeV and 28\,MeV with the magnetic spectrometer COSY-11. The new data show a significant enhancement of the total cross section compared to pure phase space expectations or calculations within a one boson exchange model. In addition, we present invariant mass spectra of two particle subsystems. While the $K^+K^-$ system is rather constant for different invariant masses, there is an enhancement in the $pK^-$ system towards lower masses which could at least be partially connected to the influence of the $\Lambda(1405)$
resonance..
\end{abstract}

\begin{keyword}
kaon \sep antikaon \sep strangeness \sep near threshold meson production
\PACS 13.60.Hb \sep 13.60.Le \sep 13.75.-n \sep 25.40.Ve  
\end{keyword}
\end{frontmatter}

\section{Introduction\label{introduction}}
The strength of the kaon-antikaon interaction appears to be very essential with respect to different physics topics. It is an important parameter in the ongoing discussion on the nature of the scalar resonances $a_0$ and $f_0$ in the mass range of $\sim\!\!1\,$GeV/c$^2$. Besides the interpretation as a $q\bar{q}$ meson \cite{morgan:93}, these resonances were also proposed to be $qq\bar{q}\bar{q}$ states \cite{jaffe:77}, $K\!\bar{K}$ molecules \cite{weinstein:90,lohse:90}, hybrid $q\bar{q}$/meson-meson systems \cite{beveren:86} or even quark-less gluonic hadrons \cite{jaffe:75}. Especially for the formation of a molecule, the strength of the $K\!\bar{K}$ interaction is a crucial quantity and it can be probed in the $K\!\bar{K}$ production close to threshold \cite{krehl:97}.\\
Due to the unavailability of kaon targets for the analysis of $K\!\bar{K}$ scattering, the kaon pair production in multi particle exit channels like \reaktion is the only possibility to study this interaction by selecting the appropriate kinematic region of the phase space distribution. Besides the $K\!\bar{K}$ subsystem, information about the $K\!N$ system is of equal importance especially in view of the actual discussion on the structure of the excited hyperon $\Lambda(1405)$ which is considered as a 3 quark system or a $K\!N$ molecular state \cite{kaiser:95}. Up to now the scattering length $a_{K^-p}$ has been mainly determined on kaonic hydrogen. But the situation is not yet clarified since first, the results of former measurements \cite{davies:79,izycki:80,bird:83,ito:98} and preliminary results at DEAR \cite{guaraldo:04,cargnelli:04} are in disagreement and second, it has been shown that contrary to pionic hydrogen, the isospin violating correction cannot be neglected in the kaonic case \cite{meissner:04}. Due to these still open questions and the fact that the analysis of former $K\!N$ data (cf. figure 22 in \cite{ito:98} and references therein) have a rather large variation, new low energy $pK^-$ scattering data can provide an independent contribution to this important issue.\\
Furthermore, a precise knowledge of the $K^\pm$ cross sections and a good understanding of the kaon and antikaon interaction with the nucleon is an essential ingredient for calculations of medium effects \cite{senger:99} being related to open questions of astrophysics \cite{brown:93}. This is because in dense matter processes there are secondary production mechanisms on hyperons such as $\pi Y \to K^-\! N$ making it necessary to understand the production above threshold \cite{roth:05}.\\
While the database on the elementary $K^+$ creation covers a wide energy range \cite{balewski:96,bilger:98,balewski:98-2,sewerin:99,kowina:04,rozek:05}, low energy data on the $K^-$ production are less available \cite{wolke:97,quentmeier:01-2,balestra:00}. In the near threshold regime, the excitation function might show a significant difference compared to the expectation of a pure phase space because final state interaction effects are predominant at low relative energies of the outgoing particles \cite{moskal:02-3}.\\
Due to the mentioned aspects together with the tendency becoming apparent that the available data seem to lie above theoretical expectations, we performed two new measurements of the total cross section of the reaction \reaktion at excess energies of $Q=10$\,MeV and $28$\,MeV \cite{winter:05} in order to further study this enhancement and its strength. In the next section, we describe the experimental technique followed by the presentation of the results. 

\section{Experiment\label{experiment}}
\begin{figure}[ht]
\centerline{\ifbw \epsfig{file=cosy11_bw,width=0.72\columnwidth}%
\else \epsfig{file=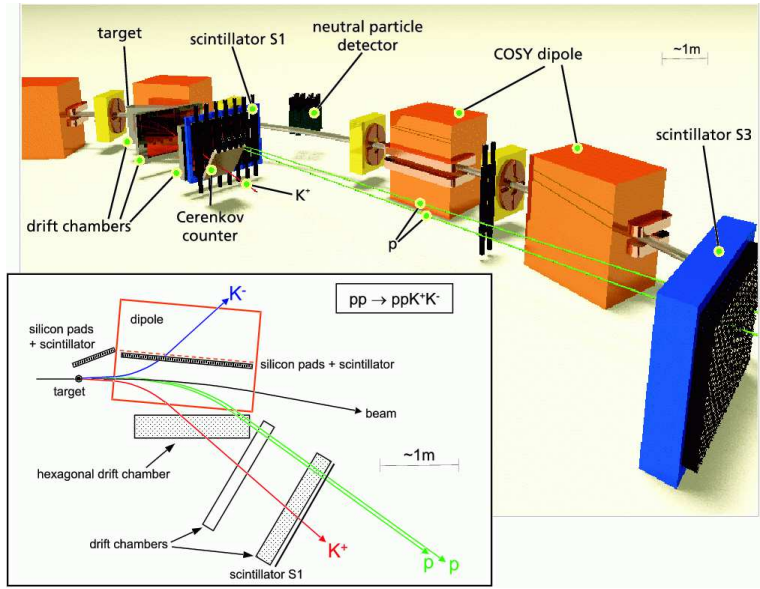,width=0.72\columnwidth}\fi}%
\caption{The experiment COSY-11 with the main components. The overlayed box shows a schematic view for an exemplary event of the reaction channel \reaktion\hspace{-1ex}. The sudden stop of the kaon track in the three dimensional picture indicates its decay.\label{cosy11}}
\end{figure}

The measurements of the \reaktion reaction were performed with the internal experiment COSY-11 \cite{brauksiepe:96} at the COoler SYnchrotron COSY \cite{maier:97nim} in J\"ulich with a beam momenta of $p=3.333$\,GeV/c and 3.390\,GeV/c corresponding to excess energies of $Q=10$\,MeV and  $28$\,MeV, respectively. Both energies are below the $\phi(1020)$ meson production threshold. The detector which is shown in figure \ref{cosy11} is designed as a magnetic spectrometer.
Using a hydrogen cluster target \cite{dombrowski:97} in front of one of the COSY dipole magnets,  positively charged particles in the exit channel are bent more -- compared to the remaining beam protons -- towards the interior of the ring where they are detected in a set of three drift chambers \cite{smyrski:05}. Tracing back the reconstructed trajectories through the known magnetic field \cite{smyrski:97} to the interaction point allows for a momentum determination. In combination with a subsequent time of flight measurement over a distance of 9.4\,m between two scintillation hodoscopes S1 and S3, these particles are identified via their invariant masses. Due to the decay of the kaon, the probability that it reaches the stop counter S3 is in the order of a few percent. Therefore, an indirect reconstruction of the time of flight is used. After the determination of the two protons' four momenta in combination with the known length of their flight path from the target to the S1 detector, the time of the interaction is calculated. This time is then used as the start for the kaon's time of flight between the point of interaction and the crossing in the S1 scintillator in order to derive the four momentum of the $K^+$.\\
Two additional detector components are mounted in front of the dipole magnet close to the target and inside the dipole gap both consisting of a scintillator and silicon pads. While the first is used to measure the coincident proton of the $pp$-elastic scattering, the array in the dipole gap serves to detect the $K^-$\!.\\
The analysis for the reaction \reaktion proceeds in several steps. First, events with less than three reconstructed tracks are rejected.
\begin{figure}[ht]
\centerline{\epsfig{file=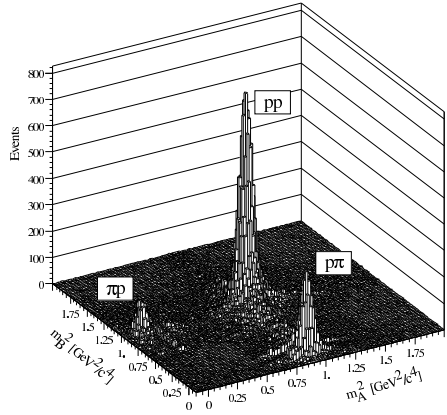,width=0.42\columnwidth}}
\caption{Identification of the two protons via the squared invariant mass for events with three reconstructed tracks. Here, the two particles are encountered which reach the final scintillator S3.\label{invmass}}
\end{figure}
For the remaining data, figure \ref{invmass} shows the squared invariant masses for those two tracks that could be assigned to a hit in the S3 scintillator. A clear separable peak for two protons is visible. With the described indirect method for the time of flight, the four momentum of the third positive particle is deduced. Figure \ref{invmassvsmissmass} shows the squared mass of the third particle $X^+$ versus the missing mass of the three particle system assuming that $X^+$ is a kaon. The z-axis is in a logarithmic scale. 
\begin{figure}[ht]
\centerline{%
\subfigure[\label{invmassvsmissmass}]{%
\ifbw \epsfig{file=invariantvsmissingppk_bw,width=0.4\columnwidth}%
\else \epsfig{file=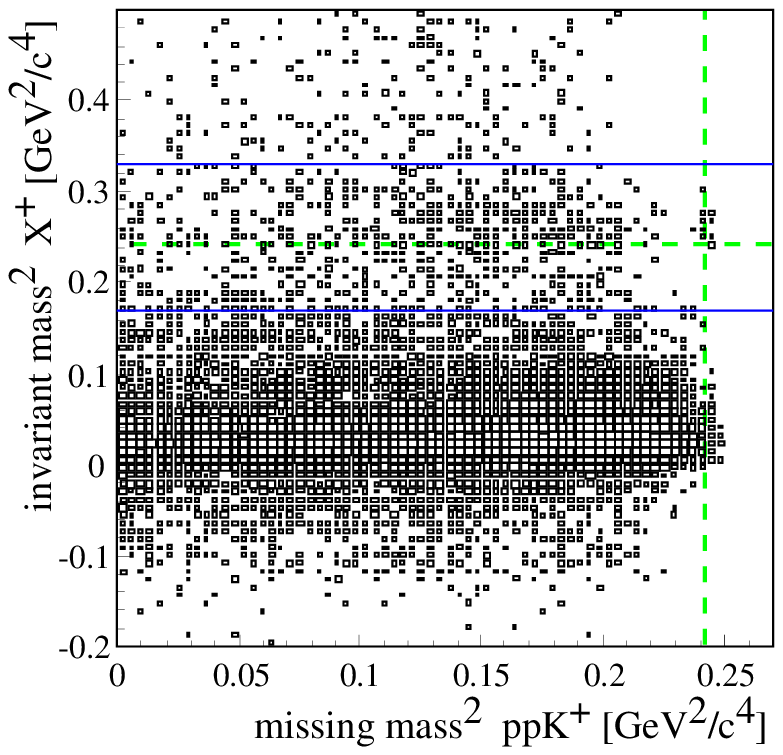,width=0.4\columnwidth}\fi}%
\hspace{0.8cm}
\subfigure[\label{ppkmissmass}]{\epsfig{file=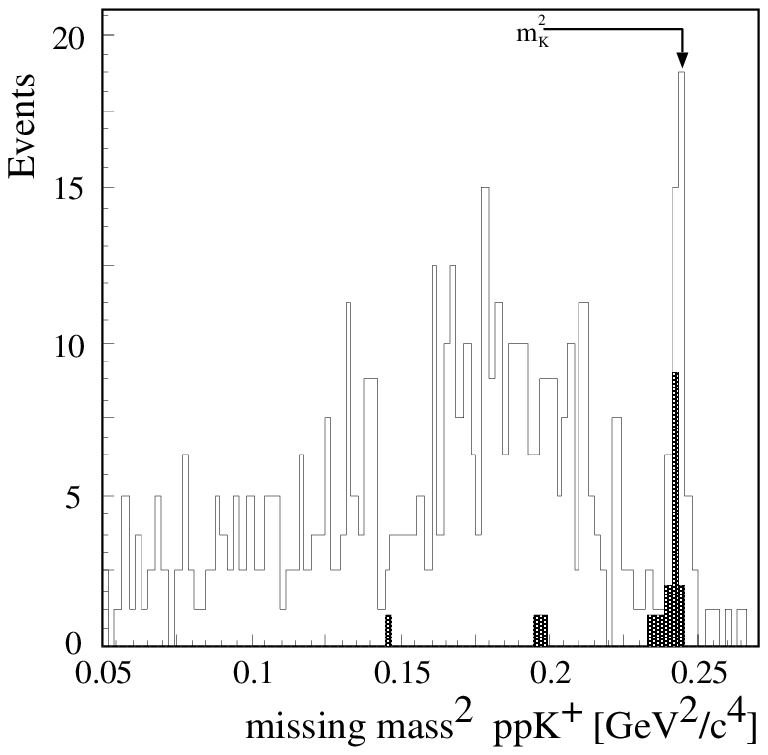,width=0.4\columnwidth}}}
\caption{a) Squared invariant mass of the third positive particle $X$ versus the missing mass assuming $m^{}_{X^+} = m^{}_{K^+}$. The \ifbw \else green \fi dashed lines indicate the literature kaon mass. The z-axis is in logarithmic scale. b) Missing mass of the $ppK^+$ system for events in between the \ifbw \else blue \fi solid lines in the left plot. An additional request for hits of certain modules in the S1 scintillator was required. The shaded spectrum includes further cuts described in the text. It is worth noting that the missing mass resolution is less than the size of the bins.}
\end{figure}
Besides the huge horizontal pionic band there is a less pronounced band structure along the kaon mass. At the cross point of both \ifbw \else green \fi dashed lines (corresponding to the literature value of the charged kaon mass \cite{eidelman:04}) a separated group of events is visible.\\
The projection within a broad band around the kaon mass (horizontal \ifbw \else blue \fi solid lines in fig. \ref{invmassvsmissmass}) is shown in figure \ref{ppkmissmass}. Here, additional cuts on the segments in the S1 scintillator were applied since the protons are passing the detector closer to the beam pipe than the $K^+$\!. These cuts were adjusted with Monte Carlo simulations for both energies separately. The missing mass spectrum (fig. \ref{ppkmissmass}) shows a clear peak at the $K^-$ mass and a broad physical background towards lower missing masses. The latter is understood mainly with the excited hyperon production $pp\to pK^+ \Lambda(1405)/\Sigma(1385)$ where the second proton stems from the decay of the resonance. Additionally, pion production channels contribute where pions are misidentified as kaons \cite{quentmeier:01-2}.\\
A fit of the kaon peak with a Gaussian and a polynomial function describing the background results in a missing mass resolution of $\sigma^{}_{MM}\approx 1$\,MeV/c$^2$ for both energies. Additional cuts using the detector mounted inside the dipole gap can be used to drastically reduce the background. The explicit procedure can be found in \cite{quentmeier:01-2,winter:05} and was performed to cross-check the signal resulting from the $K^-$ production. The principle procedure is to calculate with the known $K^-$ four momentum its expected hit position in the silicon pads inside the dipole gap and to compare it with the experimental location. While for real $K^-$ events this should be the same, a background event does not show a correlation of these two values. The final and nearly background free missing mass spectrum is shown in figure \ref{ppkmissmass} (shaded area). The loss of kaons due to its decay agrees perfectly with Monte Carlo studies. Nonetheless, the deduction of the total cross section is based on the non-shaded missing mass plotted in figure \ref{ppkmissmass} due to the higher statistics for the kaon signal. \notiz{R1 (i)}The background contribution was extracted assuming a polynomial function to describe its shape and the signal to background ratio was 5.4 and 3.8 for $Q=10\,$MeV and 28\,MeV, respectively.

\section{Results\label{results}}
\subsection{Luminosity and efficiency}
The absolute normalization of the counting rate requires both the knowledge of the luminosity $\mathcal{L}$ and the total detection efficiency $\mathcal{E}$ including the geometrical acceptance of the detector and the reconstruction efficiency. For the determination of the luminosity, the elastic proton proton scattering is used. While one proton is registered in the main detector as presented above for the reaction products and therefore its four momentum is determined, the second proton is registered in the scintillator and silicon pads in front of the dipole magnet \cite{moskal:01}. The differential counting rates are normalized to the EDDA data \cite{albers:97}. The determined integrated luminosity  $\int \mathcal{L}\,dt$ is given in table \ref{lumi}, including statistical and systematical errors.\\
The total detection efficiency for the $K^+\!K^-$ production reaction was investigated using Monte Carlo simulations based on the GEANT 3 code \cite{geant:93}. This software package has been designed to completely describe the response of the detector. It is important to mention that the COSY-11 results obtained so far at similar excess energies but in other reaction channels are in very good agreement with measurements at other laboratories. In particular, results of measurements at COSY--11 for the $pp \to pp\eta'$ reaction \cite{moskal:00-02} with a beam momentum by only 1\% lower to the one reported in the present article, are in excellent agreement with the results obtained at the SATURNE facility \cite{hibou:98}.\\
Using this software package, for each generated event a detection system response is calculated and the simulated data sample is analyzed with the same program which is used for the analysis of the experimental data. The total efficiency $\mathcal{E}$ for the free reaction \reaktion including the $pp$ final state interaction is listed in table \ref{lumi}. Here, the identification of two protons and a $K^+$ in selected segments of the S1 is required. The systematical error comprises the detection and reconstruction efficiencies, the decay of the kaon and the variance of the efficiency resulting from the uncertainty of the beam ($\Delta p/p \leq 10^{-3}$) that has been extracted to be in the order of 1-2 MeV/c \cite{winter:05}. Due to the scaling of the efficiency with $1/Q$, this translates to a variation of $\mathcal{E}$ by 9\% and 1\% at Q=10 MeV and Q=28MeV, respectively.\\
Furthermore, we included an estimate of the influence of higher partial wave contributions. From measurements at even higher excess energies \cite{quentmeier:01-2} and the extracted angular spectra \cite{winter:05} there is no indication of a substantial contribution from higher partial waves. In consequence, the additional term for the systematical error of a few percent is rather overestimating and a conservative upper estimate.\\
The \notiz{R2, 3) NB 1}inclusion of the $pp$-FSI to the Monte-Carlo studies results in a relative change of $\mathcal{E}$ by 10\% \cite{moskal:00,moskal:00-02}. Since the various descriptions for the $pp$-FSI differ from each other by only around 30\%, the contribution to the relative systematic error of $\mathcal{E}$ from the inaccuracy of the knowledge of the FSI is 3\% (30\% out of 10\%). Possible influences stemming from the $pK$ and $KK$ FSI are negligible compared to the $pp$-FSI\footnote{For the pp-FSI the scattering length $a_{pp}=7.8\,$fm \cite{naisse:77} is much larger compared to the one for the pK system  ($a_{pK}$) being less than 1\,fm.  Since the FSI is in first order proportional to the squared scattering length this neglect is reasonable.} and therefore were not included.\notiz{\rule[7cm]{0cm}{0.1cm}R1, iii)}\notiz{\rule[3cm]{0cm}{0.1cm}R1 ii)}
\begin{table}[hb]
\caption{The determined total integrated luminosity including statistical and systematical errors and the total detection efficiency for both $Q$ values.\label{lumi}}
\centerline{
\begin{tabular}{ccc}
\hline
& $Q=10$\,MeV & $Q=28$\,MeV \\
\hline
$\int \mathcal{L}\,dt\ $[pb$^{-1}$]& $2.770 \pm 0.045 \pm 0.011$& $2.270 \pm 0.064 \pm 0.006$\\
$\mathcal{E}\ [\%]$ & $1.238 \pm 0.129$ & $0.308 \pm 0.027$\\
\hline
\end{tabular}}
\end{table}

\subsection{Total cross section}
\begin{figure}[hb]
\subfigure[\label{wqall}]{\epsfig{file=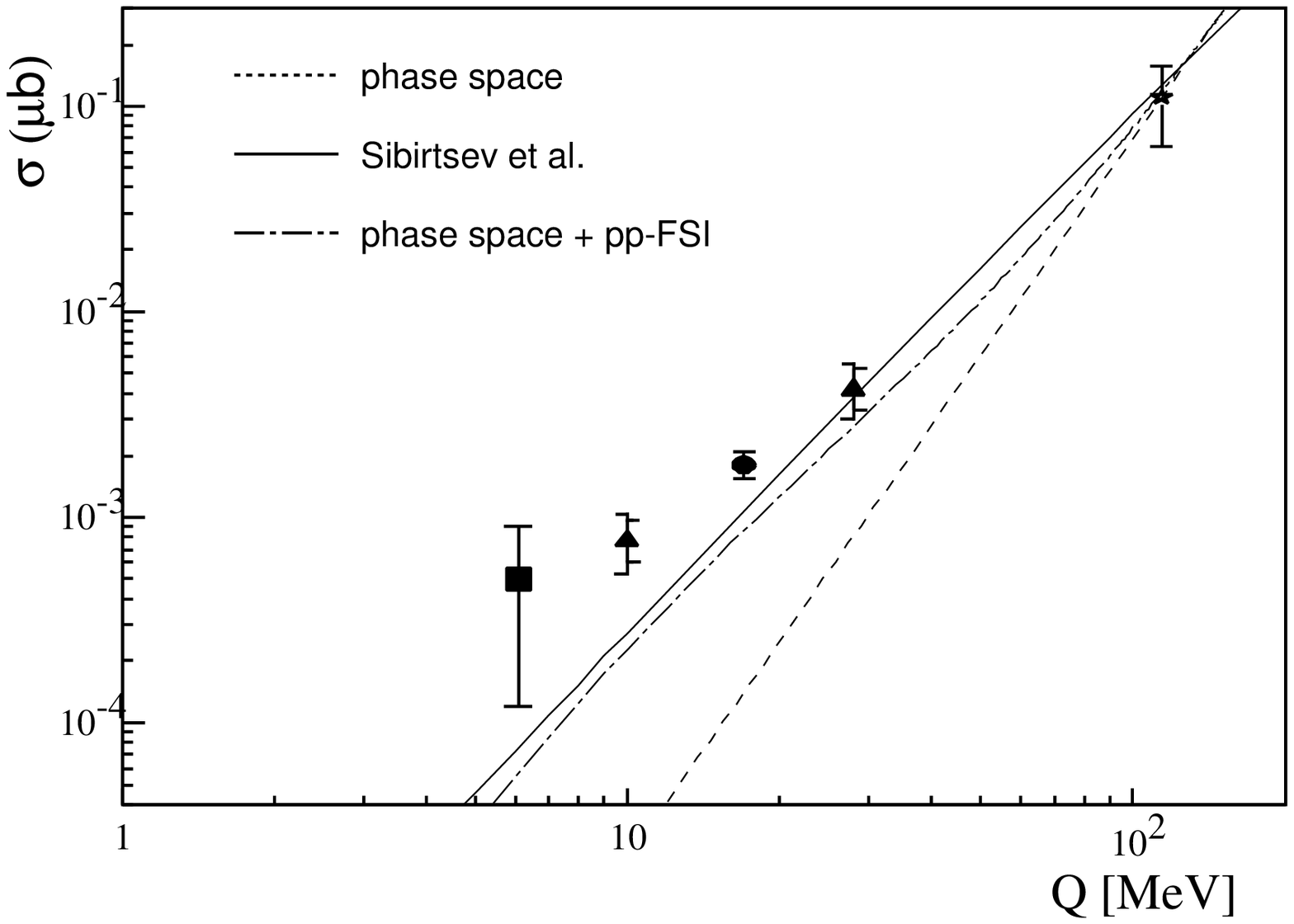,height=5.8cm}}\hfill
\subfigure[\label{ratio}]{\epsfig{file=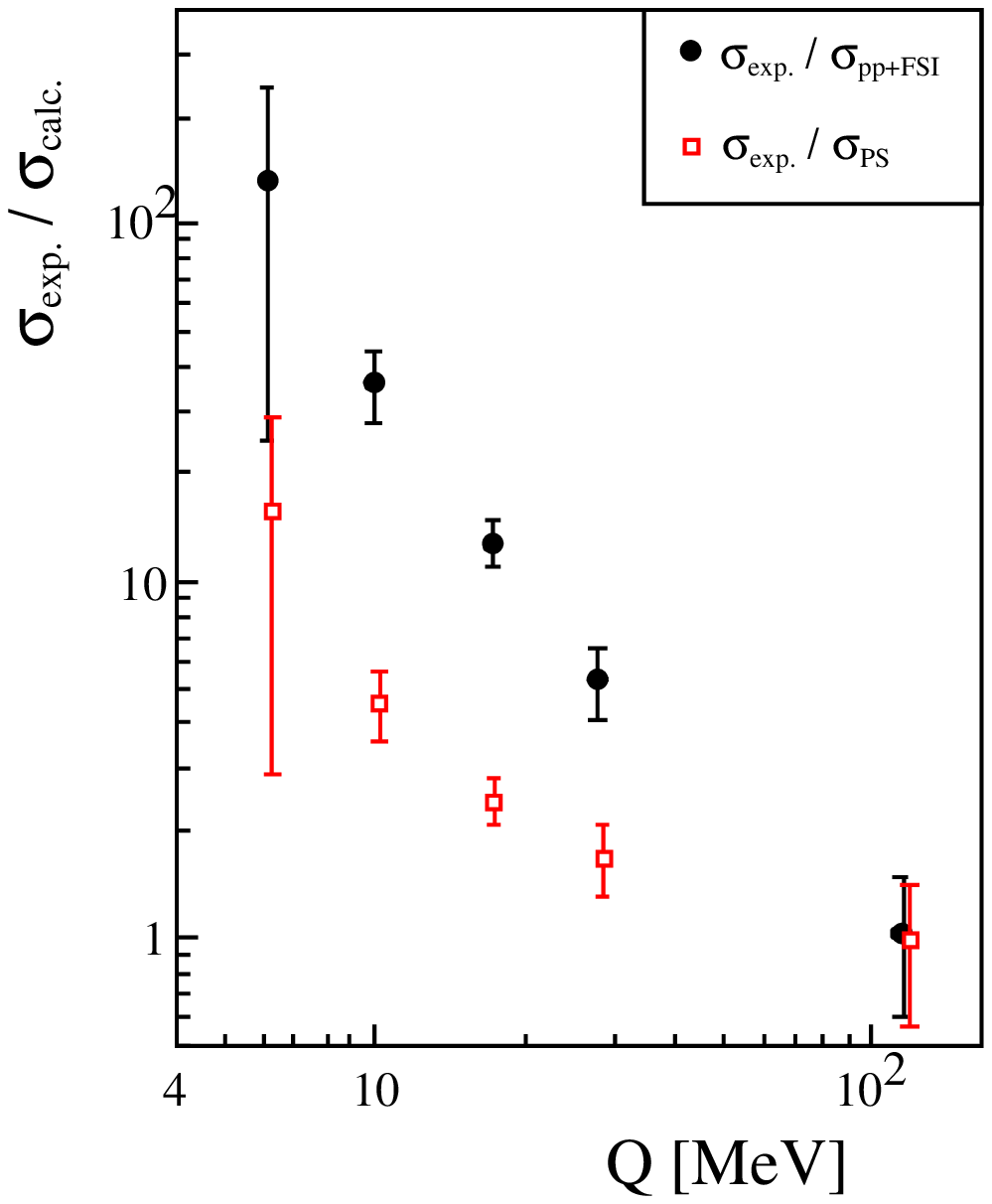,height=5.8cm}}
\caption{a) Total cross section as a function of the excess energy $Q$ for the reaction \reaktion\hspace{-1ex}. The former experimental results are taken from \cite{wolke:97,quentmeier:01-2,balestra:00} and shown together with the new results (triangles) for which data the statistical error (horizontal right markers) and the statistical plus systematic errors (horizontal left markers) are shown. b) Ratio of the experimental data $\sigma_{exp}$ over the calculation for pure phase space and for the parametrization of the $pp$-FSI.}
\end{figure}

Using the knowledge of the integrated luminosity and the overall efficiency, the number of events registered for both excess energies can be transformed into a total cross section $\sigma_{tot}$. For both energies, we obtained the final results given in table \ref{cross},
\begin{table}[ht]
\caption{Total cross section for both $Q$ values including statistical and systematical errors.\label{cross}}
\centerline{
\begin{tabular}{cc}
\hline
$Q$\,[MeV] & $\sigma_{tot}$ [nb] \\
\hline
10 & $\boldsymbol{0.787 \pm 0.178 \pm 0.082}$\\
28 & $\boldsymbol{4.285 \pm 0.977 \pm 0.374}$\\
\hline
\end{tabular}}
\end{table}
where the contributing systematical errors were added quadratically. The new results together with previous measurements are compiled in figure \ref{wqall} together with some theoretical expectations.\\
It is obvious that at low excess energies the data points lie significantly above the expectations indicated by the different lines that are all normalized to the DISTO point at $Q=114\,$MeV. \notiz{\rule[2cm]{0cm}{0.1cm}R2, 4) and Pawel\\$\vdots$}The reason for this choice is the fact that the effect of the nucleon-nucleon FSI diminishes with increasing excess energy since it significantly influences only that part of the phase space at which nucleons have small relative momenta. While this fraction stays constant, the full phase space volume $V_{PS}$ grows rapidly: A change from $Q=1\,$MeV to $Q=10\,$MeV corresponds to a growth of $V_{PS}$ by more than three orders of magnitude. As a result the S-wave $pp$-FSI is of less importance for higher excess energies where it affects a small fraction of the available phase space volume only \cite{moskal:04-3}. Additionally, a possible contribution of higher partial waves at $Q=114\,$MeV should even increase the total cross section at this energy. Therefore, their inclusion in the calculations would result in an even stronger discrepancy between the calculations and the data at low $Q$ values.\\
The pure non-relativistic phase space (dashed line in fig. \ref{wqall}) differs from the experimental data by two orders of magnitude at $Q=10\,$MeV and a factor of five to ten at $Q=28\,$MeV. In comparison to that, the inclusion of the $pp$-FSI (\ifbw dashed-dotted \else red solid \fi line) by folding its parameterization known from the three body final state with the four body phase space is clearly closer to the experimental results but does not fully account for the difference\footnote{The parametrization presented differs from that in figure 6 of reference \cite{quentmeier:01-2}. The reason is that the formula used in \cite{quentmeier:01-2} was originally derived for the three body final state. Our new ansatz \cite{winter:05} avoids this approximation by splitting the integral over the four body phase space into two parts whereas one of them contains a three body subsystem  for which the pp-FSI parametrization is known.}\notiz{\rule{0cm}{0.1cm}\\\rule[1cm]{0cm}{0.1cm}R2, 4)}\notiz{\rule{0cm}{0.1cm}\\\rule[5cm]{0cm}{0.1cm}R1, iv) and v)}. The solid line representing the calculation within a one-boson exchange model \cite{sibirtsev:97} reveals a similar discrepancy as the $pp$-FSI parameterization. This model includes an energy dependent scattering amplitude derived from the fit of the total cross sections in $K^\pm p \to K^\pm p$ \cite{baldini:88} while the $pp$-FSI was not included, yet. Up to now, there is no full calculation available but the new data demand further theoretical efforts in order to give a complete picture of the $K^+\!K^-$ production.\\
The enhancement of the total cross section at low energies could be partly induced by the opening of the neutral ($K^0\!\bar{K}^0$) kaon pair channel with a mass splitting compared to the charged ($K^+\! K^-$) kaon pairs of about 8\,MeV. Since calculations show a substantial influence of the opening neutral channel on the $\pi\pi \to K^+\!K^-$ cross section (cf. figure 2 in reference \cite{krehl:97}), an effect on the excitation function of the reaction \reaktion could be expected at low energies as well. A quantitative estimation of this effect has to be calculated within a complete coupled channel model taking into account this $K^+\!K^- \!\rightleftharpoons\! K^0\!\bar{K}^0$ transition. However, the kinematical situation of the four body final state is expected to strongly suppress the effect \cite{haidenbauer:05} compared to the two body final state $\pi\pi \to K^+\!K^-$.

\subsection{Differential spectra}
\begin{figure}[ht]
\centerline{%
\subfigure[\label{kpkm}]{\epsfig{file=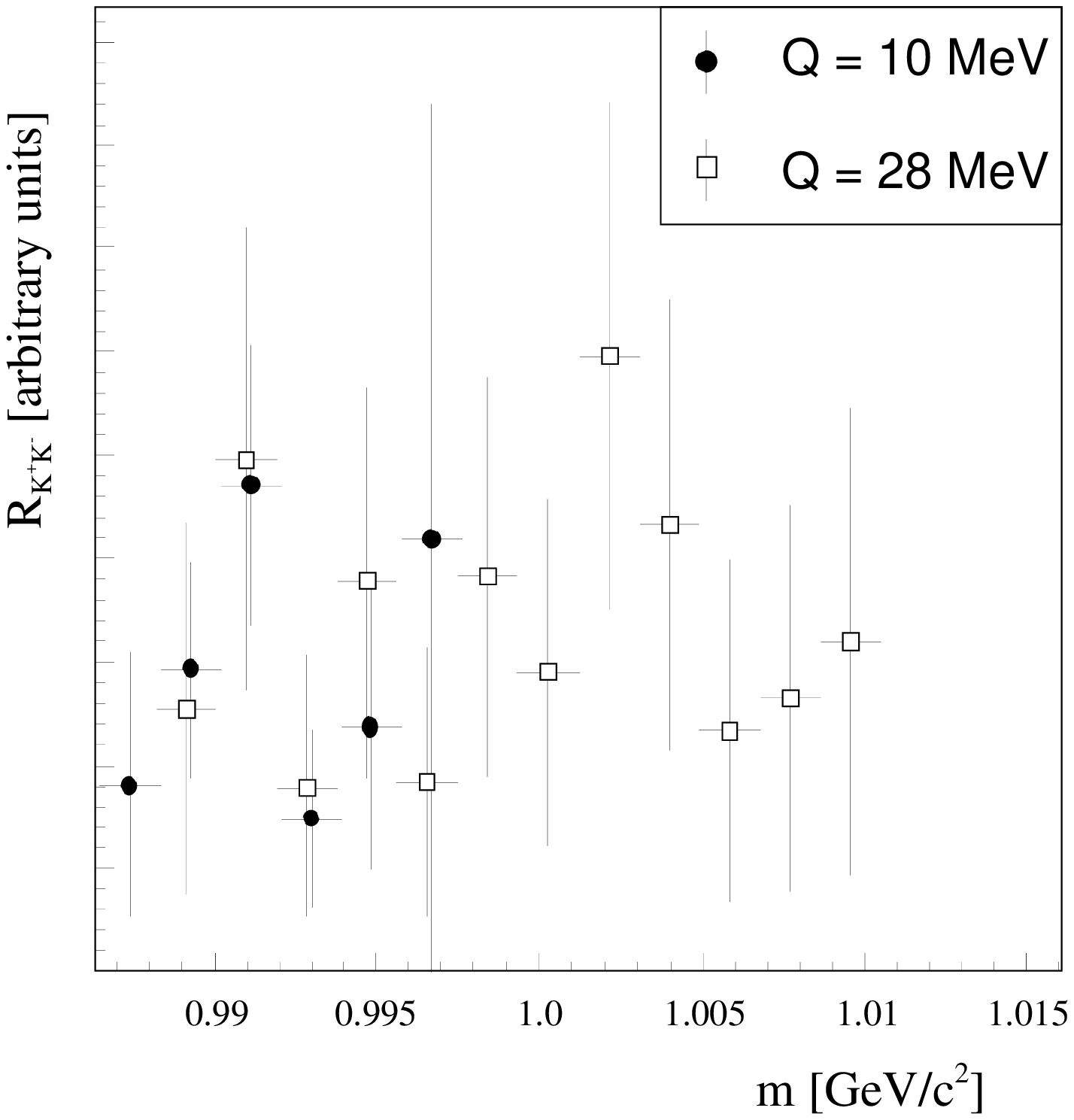,height=0.41\columnwidth}}
\hspace{0.8cm}
\subfigure[\label{pkmpkp}]{\epsfig{file=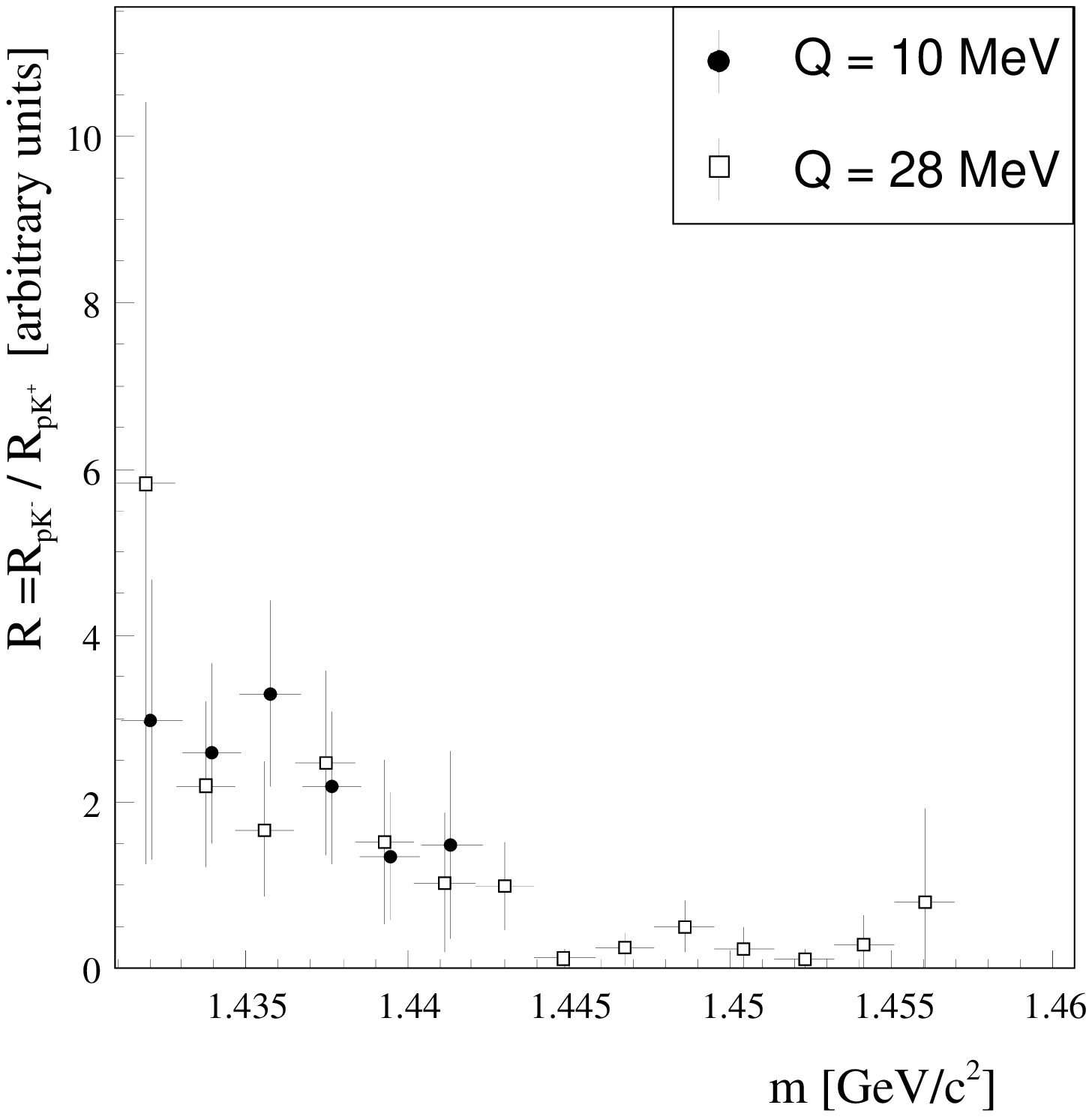,height=0.41\columnwidth}}}
\caption{a) Invariant mass $m_{K^+\!K^-}^{}$ for both excess energies normalised to the accepted MC events generated with a phase space distribution. b) Invariant mass for the system $pK^-$ divided by that for the $pK^+$ system. For a better clearness the data points at $Q=10\,$MeV were slightly shifted right for both pictures.}
\end{figure}

The knowledge of the four momenta allows to investigate differential observables. Here, in particular the invariant mass of several subsystems can be exploited. For two particles $i$ and $j$ in the exit channel, the square of the invariant mass\footnote{In case of $m_{pK^+}^{}$ or $m_{pK^-}^{}$, for each event both protons are taken into account so that each invariant mass $m_{pK}^{}$ has two entries per event.} $m_{ij}^2$ is given by $m^2_{ij}=(\mathbb{P}_i+\mathbb{P}_j)^2$ with $\mathbb{P}_i$ being the four momentum of the particle $i$. The events for the following plots were chosen under the condition, that the missing mass $m_X^{}$ in figure \ref{ppkmissmass} is within the region of the kaon (0.235\,GeV$^2$/c$^4 < m_X^2 < 0.25$\,GeV$^2$/c$^4$). Furthermore, the events were corrected by the detection efficiency and then normalised to the phase space distribution\footnote{Therefore, instead of the mass spectrum $m_{ij}$ we show the ratio $R_{ij}^{} := \frac{dN}{dm_{ij}^{}} / \frac{dN}{dm_{ij}^{PS}}$, where $\frac{dN}{dm_{ij}^{}}$ is the number of events registered in a certain mass bin of the subsystem $ij$ and $\frac{dN}{dm_{ij}^{PS}}$ is the number of accepted Monte Carlo events in the same mass bin generated with an underlying phase space distribution. Deviations from the phase space directly reflect into a non flat distribution of $R_{ij}$.}. Generally for each mass bin a separate background subtraction should be performed in order to make a reliable interpretation. For high statistics this in fact has been done (e.g. in case of the $pp \to pp\eta$ \cite{moskal:04-2}). For the present data, however, the background contribution is that small (as seen in figure~\ref{ppkmissmass}), that such a procedure could be avoided.\\
For a pure phase space distribution, the ratio $R_{K^+K^-}^{}$ should be flat as it rather is in case of the $K^+\!K^-$ system shown in figure \ref{kpkm} for both $Q$ values. \notiz{R1, vi)}In a former publication \cite{quentmeier:01-2} it has been shown, that a possible influence of the scalar resonances $a_0(980) / f_0(980)$ is not distinguishable from a pure $s$-wave distribution within the current statistics.\\
Figure \ref{pkmpkp} shows the double ratio $R := R_{pK^-}^{} / R_{pK^+}^{}$. The normalisation of the $pK^-$ system to the $pK^+$ has the advantage, that any systematic errors from misunderstood inefficiency should in principle affect both systems in a similar way and therefore mainly cancel out. In consequence, since the interaction in the $pK^+$ system is known to be rather weak\footnote{The experimental distribution for this subsystem is indeed within the error bars not deviating from the pure phase space (similar like in the case of the $K^+K^-$ system in figure \ref{kpkm}).}\notiz{\rule[4cm]{0cm}{0.1cm}\\\rule[6cm]{0cm}{0.1cm}R2, 2)}, any non flat distribution in this ratio $R$ might better indicate an interaction between the proton and the negative kaon than the pure spectrum of $m_{pK^-}^{}$ alone. There is a clear increase towards lower invariant masses in the double ratio $R$. This could result from the final state interaction of the two particles or partially be a reflection of the $\Lambda(1405)$\footnote{The $\Sigma(1385)$ resonance certainly will influence this system as well. However, due to it's lower mass and slightly smaller width \cite{eidelman:04}, it's contribution should be less significant than that of the $\Lambda(1405)$.} or a mixture of both.\\
The shown distributions might trigger further investigations with improved statistics with the new installation of the WASA detector \cite{zabierowski:02,adam:04} which combines a large acceptance of nearly $4\pi$ with the simultaneous detection of charged and neutral particles.\notiz{\rule[2cm]{0cm}{0.1cm}\\\rule[3cm]{0cm}{0.1cm}R2, 2)}

\section{Summary}
The COSY-11 collaboration has extended its studies on the elementary $K^-$ production by measuring the reaction \reaktion at excess energies of $Q=10$\,MeV and 28\,MeV resulting in total cross sections exceeding the expectations for a pure phase space drastically. The measurement is based on a kinematically complete reconstruction of the positively charged ejectiles while the negative kaon is identified via the missing mass. An absolute normalization of the counting rate is achieved via a simultaneous measurement of the $pp$ elastic scattering.\\
The new results for the total cross section in the reaction \reaktion are clearly showing that towards the lower $Q$ values the data are exceeding any expectations both from pure phase space with and without the $pp$ FSI and a calculation within a meson exchange model. To further study the strength of the enhancement of the total cross section at low $Q$ values, the COSY-11 collaboration will remeasure the data point at $Q=6$\,MeV \cite{winter:04-4} in order to significantly reduce the statistical error.\\
Within the limited statistics the differential $m_{pK^-}^{}$ distribution normalised to the $pK^+$ system shows a $pK^-$ interaction which might have a connection to the $\Lambda(1405)$. \notiz{R2, 1) and 2)}It is too early to extract quantitative coupling strengths information from the present data on the hyperon resonances $\Sigma(1385)$ and $\Lambda(1405)$ where especially the structure of the latter one is under discussion. But the present data clearly demonstrate the sensitivity in this $ppK^+K^-$ final state to the $KN$ interaction which is an important ingredient in the interaction of the $\Lambda(1405)$ as a bound $KN$ system.

\section{Acknowledgments}
We would like to thank J. Haidenbauer, C. Hanhart, and A. Sibirtsev for helpful discussions on theoretical questions. This work has been supported by the European Community - Access to Research Infrastructure action of the Improving Human Potential Programme, by the FFE grants (41266606 and 41266654) from the Research Center J{\"u}lich, by the DAAD Exchange Programme (PPP-Polen), by the Polish State Committee for Scientific Research (grant No. PB1060/P03/2004/26), and by the RII3/CT/2004/506078 - Hadron Physics-Activity -N4:EtaMesonNet.

\bibliography{abbrev,polarisation,general}
\end{document}